\documentclass[technote]{IEEEtran}
\ifCLASSINFOpdf
\else
\fi
\usepackage{xfrac}
\usepackage{amsmath}
\usepackage{multirow}
\usepackage{color}
\usepackage{amssymb}
\usepackage{cite}
\DeclareMathOperator*{\argmin}{argmin}
\usepackage{hyperref}
\usepackage{lipsum}
\usepackage{soul}
\usepackage[para]{threeparttable}
\usepackage{algpseudocode,algorithm}
\usepackage[usestackEOL]{stackengine}
\ifCLASSOPTIONcompsoc
\usepackage[caption=false,font=normalsize,labelfon
t=sf,textfont=sf]{subfig}
\else
\usepackage[caption=false,font=footnotesize]{subfi
g}
\fi

 \usepackage{etoolbox}

\usepackage{stfloats}
\usepackage{etoolbox}

 \makeatletter
\patchcmd{\@maketitle}
  {\addvspace{0.5\baselineskip}\egroup}
  {\addvspace{-.2\baselineskip}\egroup}
  {}
  {}
\makeatother

\begin{document}
\bstctlcite{IEEEexample:BSTcontrol}

%
\twocolumn{ \title{ A Lightweight Machine Learning Assisted Power Optimization for Minimum Error in NOMA-CRS over Nakagami-$m$ channels}}

%
%
%


\author{ Ferdi Kara,~\IEEEmembership{Senior Member,~IEEE,} Hakan Kaya, Halim Yanikomeroglu,~\IEEEmembership{Fellow,~IEEE.}
 \thanks{The work of F. Kara and H. Kaya is supported by Zonguldak Bulent Ecevit University with the project number 2021-75737790-02. F. Kara and H. Kaya are 
with the Electrical-Electronics Engineering, Zonguldak Bulent Ecevit University, Zonguldak, Turkey,  e-mail: \{f.kara, hakan.kaya\}@beun.edu.tr. F. Kara and H. Yanikomeroglu are with the Department of Systems and Computer Engineering, Carleton University, Ottawa, K1S 5B6, ON, Canada, e-mail:halim@sce.carleton.ca.}}

\maketitle
\begin{abstract}
Non-orthogonal multiple access based cooperative relaying system (NOMA-CRS) has been proposed to alleviate the decay in spectral efficiency of the conventional CRS. However, existing NOMA-CRS studies assume perfect successive interference canceler at the relay and mostly investigate sum rate whereas the error performance has not been taken into consideration. In this paper, we analyze error performance of the NOMA-CRS and the closed-form bit error probability (BEP) expression is derived over Nakagami-m fading channels. Then, thanks to the high performance of machine learning (ML) in challenging optimization problems, a joint power sharing-power allocation (PS-PA) scheme is proposed to minimize the bit error rate (BER) of the NOMA-CRS. The proposed ML-assisted optimization has a very low online implementation complexity. Based on provided extensive simulations, theoretical BEP analysis is validated. Besides, the proposed ML-aided PS-PA provides minimum BER (MBER) and outperforms previous PA strategies for the NOMA-CRS notably.
\end{abstract}
\begin{IEEEkeywords}
error analysis, cooperative relaying, NOMA, optimum power allocation, machine learning, Nakagami-m fading
\end{IEEEkeywords}

\IEEEpeerreviewmaketitle
\vspace{-1.5\baselineskip}
\section{Introduction}
Non-orthogonal multiple access (NOMA) has been envisioned as a key technology for the future wireless networks due to its high spectral efficiency \cite{Vaezi2019}; therefore, its integration into other physical techniques has been widely investigated such as free-space optics\cite{Saxena20} and space time-block coded schemes \cite{noma_stbc}.
NOMA-based cooperative relaying system (CRS) is one of the most attracted topics since the spectral inefficiency in conventional CRS can be eliminated thanks to NOMA integration \cite{Kim2015}. The ergodic capacity of the NOMA-CRS is analyzed and it is shown that NOMA-CRS has a better capacity performance than the conventional CRS over different fading channels \cite{Kim2015,Jiao2017}. This performance enhancement in capacity has led researches to investigate NOMA-CRS and various NOMA-CRS schemes have been proposed/analyzed in terms of capacity and outage performances \cite{Xu2016,Zhang2018,Wan2019,Abbasi2019}. \textit{Xu et. al} \cite{Xu2016} propose a novel receiver design and prove that the capacity gain of the NOMA-CRS can be further improved. Then, \textit{Zhang et. al} \cite{Zhang2018} have analyzed the capacity and outage performances of the NOMA-CRS for different transmission strategies under imperfect channel state information (CSI). Then, it is assumed to be two relays in the network  and ergodic capacity is investigated \cite{Wan2019}. Moreover, \textit{Abbasi et. al} \cite{Abbasi2019} consider an amplify-forward (AF) relay in NOMA-CRS and provide an approximate expression for ergodic rate. However, existing studies mostly assume that perfect successive interference canceler (SIC) is implemented at the relay which is quite strict/unreasonable assumption and should be relaxed due to the nature of wireless communications. Besides, all in previous works, the analysis is based on only the SINR definitions which do not represent the performance when an actual modulator and/or demodulator (e.g., SIC) are implemented. In addition, in all previous studies \cite{Kim2015,Jiao2017,Xu2016,Zhang2018,Wan2019,Abbasi2019}, NOMA-CRS has been analyzed only in terms of informational-theoretic perspectives (i.e., capacity and outage) whereas only in \cite{Li2019,Kara2019}, the approximate bit error probability (BEP)  has been conducted for only two subsets of the NOMA-CRS. Besides, these papers \cite{Li2019,Kara2019} consider only Rayleigh fading channels. However, to the best of the authors knowledge, the exact BEP of the NOMA-CRS has not been derived, yet, although, it is one of the most important key performance indicators (KPIs).

On the other hand, machine learning (ML) techniques have been proved to be efficient alternatives in solving challenging wireless communications problems and have started to attract great recent attention from the communication society \cite {Zappone2019}. Thus, in this paper, we take the advantage of ML in power optimization to minimize the error performance of the NOMA-CRS.
The main contributions are as follow: \begin{itemize} \item{We derive the exact BEP of the NOMA-CRS in closed-form and to the best of the authors' knowledge, this is the first study which investigates the error performance of the NOMA-CRS with the imperfect SIC, a realistic scenario.} In addition, this paper considers Nakagami-m fading channels which represent more comprehensive channel conditions than Rayleigh fading channels. Theoretical analysis is  validated via computer simulations. 
\item{We propose a lightweight machine learning (ML)-aided joint power sharing-power allocation (PS-PA) optimization for the NOMA-CRS under the minimum bit error rate (MBER) constraint. This novel solution proposes an optimization not only the first in terms of error performance but also the first joint optimization in terms of any performance metric.} Based on extensive simulations, proposed ML-aided PS-PA has performed well in predicting actual optimal values, obtained by an exhaustive search, and it outperforms previous PA strategies in terms of error performance with a very low complexity.\end{itemize}

The rest of the paper is organized as follows. In Section II, NOMA-CRS is defined. Then, Section III provides the theoretical BEP analysis. ML-aided optimum PS-PA is introduced in Section IV. In Section V, simulation results are presented. Finally, Section VI concludes the paper with discussions.
\section{System Model}
A NOMA-CRS where a source (S) is willing to reach out the destination (D) and a half-duplex relay (R) helps for it \cite{Kim2015,Jiao2017,Zhang2018} is considered. All nodes are assumed to be equipped with single antenna and the flat fading channel coefficient between each nodes (i.e., $h_\lambda $,  $\lambda=sr,sd,rd$) follows Nakagami-m distribution with shape $m_\lambda$ and spread $\Omega_\lambda$ parameters. In order to overcome the inefficiency of the conventional CRS of device-to-device communication, NOMA is applied for two intended/consecutive symbols of the destination in the first phase of the communication. Then, this total superposition-coded symbol is conveyed to both the destination and the relay, hence the received signals in the first phase are given as
\begin{equation}
y_{\lambda}=\sqrt{P_s}\left(\sqrt{\alpha}x_1+\sqrt{\left(1-\alpha\right)}x_2\right)h_{\lambda}+w_{\lambda},\ \lambda=sr,sd,
\end{equation}
where $P_s$ is the transmit power of the source. $\alpha$ is the power allocation (PA) coefficient. $x_1$ and $x_2$ are the two symbols of the destination to be transmitted simultaneously in NOMA-CRS -they are transmitted sequentially in conventional CRS-. $w_{\lambda}$ denotes the additive white Gaussian noise (AWGN) with $N_0$ variance. $\alpha<0.5$ is assumed.  Thus, in the first phase, both relay and destination detect $x_2$ symbols by treating $x_1$ symbols as noise. Then, the relay implements SIC to detect $x_1$ symbols and forwards detected $\hat{x}_1$ symbols to the destination in the second phase \cite{Kim2015,Jiao2017,Zhang2018}. The received signal by the destination in the second phase is given as
\begin{equation}
y_{rd}=\sqrt{P_r}\hat{x}_1h_{srd}+w_{rd},
\end{equation}
where $P_r$ is the transmit power of the relay. Finally, the destination detects $x_1$ symbols based on $y_{rd}$.
\section{Bit Error Probability (BEP) Analysis}
In order to derive total BEP of the NOMA-CRS, BEPs for two symbols should firstly be derived and averaged. Thus, the average BEP (ABEP) of the NOMA-CRS is given by
\begin{equation}
P^{(e2e)}\left(e\right)=\frac{P_{x_1}\left(e\right)+P_{x_2}\left(e\right)}{2},
\end{equation}
where $P_{x_1}\left(e\right)$ and $P_{x_2}\left(e\right)$ denote the BEPs of $x_1$ and $x_2$ symbols, respectively.

Since the $x_2$ symbols are conveyed to the destination only in the first phase and the $x_1$ symbols are treated as noise in detection, the BEP for $x_2$ symbols will be the same with BEP of \textit{far user} in downlink NOMA. The conditional BEP of \textit{far user's} symbols in NOMA schemes is given as
\begin{equation}
P_{x_2}(e|_{\gamma_{sd}})=\sum_{i=1}^N\varsigma_iQ\left(\sqrt{2\nu_i\rho_s\gamma_{sd}}\right),
\end{equation}
where $\gamma_\lambda=|h_\lambda|^2$ and $\rho_s=\sfrac{P_s}{N_0}$ are defined. $N$, $\varsigma_i$ and $\nu_i$ coefficients change according to chosen modulation constellation pairs for $x_1$ and $x_2$ symbols. In case BPSK\footnote{Most studies in the literature consider BPSK for error analysis of CRS.} is used for both symbols (i.e., $x_1$ and $x_2$), by following steps \cite[Eq.(3)-(4)]{Kara2019}, it is derived that  $N=2$, $\varsigma_i=0.5$ and $\nu_i= 1\mp2\sqrt{\alpha-\alpha^2}$.
\begin{IEEEproof}
In NOMA schemes, the BEP highly depends on chosen constellation pairs and we should consider the signal energy in each scenario (i.e., that is not fixed due to superimposed symbols). Considering the baseband symbol $x_i=\mp1$ for BPSK, the received signal at the nodes are given in Table I along with the correct decision rule. For instance, when $b_1b_2=00$ bit-stream is conveyed for $x_1$ and $x_2$ symbols, the received signal at the nodes becomes $y_\lambda=\left(-\sqrt{\alpha}-\sqrt{\left(1-\alpha\right)}\right)h_\lambda+n_\lambda$. To detect $x_2$ symbols correctly, the received signal, $y_\lambda<0$ should be satisfied; thus; the erroneous detection probability is given by $P\left(n_\lambda\geq\left(\sqrt{\alpha}+\sqrt{\left(1-\alpha\right)}\right)h_\lambda\right)$. By repeating all scenarios with the given decision rules, we obtain coefficients as given above, so the proof is completed.
\begin{table}[]
\begin{threeparttable}
    \centering
    \caption{Superimposed Symbols and Correct Decision Rules}
    \begin{tabular}{|c|c|c|}\hline
       Bit-stream ($b_1b_2$) & Received (Remained) Signal\tnote{a} & Correct Decision rule \\ \hline
         00& $-\sqrt{\alpha}-\sqrt{\left(1-\alpha\right)}$& $y_\lambda<0$  \\ \hline
         01&$-\sqrt{\alpha}+\sqrt{\left(1-\alpha\right)}$ & $y_\lambda\geq 0$  \\ \hline
         10& $\sqrt{\alpha}-\sqrt{\left(1-\alpha\right)}$& $y_\lambda<0$  \\ \hline
         11&$\sqrt{\alpha}+\sqrt{\left(1-\alpha\right)}$ & $y_\lambda\geq 0$ \\ \hline
         \multicolumn{3}{|c|}{After Correct SIC (Decoding $x_2$)} \\ \hline
          00&$-\sqrt{\alpha}$ & $y_{sr}^{'}<0$\\ \hline
         01&$-\sqrt{\alpha}$  & $y_{sr}^{'}<0$ \\ \hline
         10& $\sqrt{\alpha}$& $y_{sr}^{'}\geq0$\\ \hline
         11&$\sqrt{\alpha}$ & $y_{sr}^{'}\geq0$\\ \hline
    \multicolumn{3}{|c|}{After Erroneous SIC (Decoding $x_2$)} \\ \hline
    00& $-\sqrt{\alpha}-2\sqrt{\left(1-\alpha\right)}$& $y_{sr}^{'}<0$   \\ \hline
         01&$-\sqrt{\alpha}+2\sqrt{\left(1-\alpha\right)}$  & $y_{sr}^{'}<0$  \\ \hline
         10& $\sqrt{\alpha}-2\sqrt{\left(1-\alpha\right)}$ & $y_{sr}^{'}\geq0$ \\ \hline
         11&$\sqrt{\alpha}+2\sqrt{\left(1-\alpha\right)}$ & $y_{sr}^{'}\geq0$  \\ \hline
    \end{tabular}
\begin{tablenotes}[normal,flushleft]
\item[a]\footnotesize{For the representation simplicity, we do not represent channel coefficient and AWGN.}
\end{tablenotes}
\end{threeparttable}
    \label{tab:my_label}
\end{table}
\end{IEEEproof}
Then, by averaging the conditional BEP over instantaneous $\gamma_{sd}$ (it follows Gamma distribution ),with the aid of \cite{Alouini1999}, the BEP of $x_2$ symbols is derived as
\begin{equation*}
\begin{split}
  &P_{x_2}(e)=\\
  &\begin{cases}
  \sum\limits_{k=1}^2\frac{1}{4}\left[1-\mu^2(b_k)\sum\limits_{l=0}^{m_{sd}-1}\binom{2l}{l}\left(\frac{1-\mu^2(b_k)}{4}\right)^l\right], m_{sd}:  \text{integer}, \\
  \sum\limits_{k=1}^2\frac{1}{4\sqrt{\pi}}\frac{\sqrt{b_k}}{\left(1+b_k\right)^{m_{sd}+0.5}}\frac{\Gamma(m_{sd}+0.5)}{\Gamma(m_{sd}+1)} \times\\
  {_2}F_1\left(1,m_{sd}+0.5;m_{sd}+1;\frac{1}{1+b_k}\right),  m_{sd}: \text{non-integer},
    \end{cases}
    \end{split}\tag5
 \end{equation*} \stepcounter{equation}
where $b_k\triangleq\frac{\nu_k\rho_s\Omega_{sd}}{m_{sd}}$ and $\mu(z)\triangleq\sqrt{\frac{z}{1+z}}$ are defined. In (5), $\Gamma(.)$ and $ {_2}F_1\left(,;;\right)$ denote Gamma \cite[Eq. (8.31)]{Gradshteyn1994} and Gauss Hyper-geometric \cite[Eq. (9.10)]{Gradshteyn1994} functions, respectively.

On the other hand, the $x_1$ symbols are detected at the relay in the first phase and forwarded to the destination in the second phase. Since the erroneous detection in two phases are statistically independent, with the law of total probability, the BEP of $x_1$ symbols is given as
\begin{equation}
P_{x_1}(e)=P_{x_1}^{(sr)}(e)\left(1-P_{x_1}^{(rd)}(e)\right)+\left(1-P_{x_1}^{(sr)}(e)\right) P_{x_1}^{(rd)}(e),
\end{equation}
where $P_{x_1}^{(sr)}(e)$ and $P_{x_1}^{(rd)}(e)$ denote the BEPs of $x_1$ symbols between nodes S-R (i.e., first phase) and R-D (i.e., second phase), respectively.

The BEP of $x_1$ in the second phase  (i.e., $P_{x_1}^{(rd)}(e)$) can easily be obtained, since no interference is encountered in the second phase (only transmission from R-D exists). Thus, the conditional BEP of $x_1$ in the second phase turns out to be the well-known error probability over fading channels. For BPSK, it is given as $P_{x_1}^{(rd)}(e|_{\gamma_{rd}})=Q(\sqrt{2\rho_r\gamma_{rd}})$ where $\rho_r=\sfrac{P_r}{N_0}$. The BEP over Nakagami-m fading channels is given as \cite{Alouini1999}
\begin{equation*}
\begin{split}
  &P_{x_1}^{(rd)}(e)=\\
  &\begin{cases}
\frac{1}{2}\left[1-\mu^2(p)\sum\limits_{l=0}^{m_{rd}-1}\binom{2l}{l}\left(\frac{1-\mu^2(p)}{4}\right)^l\right], m_{rd}:  \text{integer}, \\
 \frac{1}{2\sqrt{\pi}}\frac{\sqrt{p}}{\left(1+p\right)^{m_{rd}+0.5}}\frac{\Gamma(m_{rd}+0.5)}{\Gamma(m_{rd}+1)} \times\\
  {_2}F_1\left(1,m_{rd}+0.5;m_{rd}+1;\frac{1}{1+p}\right),  m_{rd}:,  \text{non-integer},
    \end{cases}
    \end{split}\tag7
 \end{equation*} \stepcounter{equation}
where $p\triangleq\frac{\rho_r\Omega_{rd}}{m_{rd}}$ is defined.

However, in order to derive the BEP in the first phase (i.e., $P_{x_1}^{(sr)}(e)$), much more effort is required. In the first phase, since the superimposed signal is received by the relay, the SIC should be implemented and the error propagation during SIC should be taken into consideration in the BEP analysis.
The conditional BEP of $x_1$ symbols (i.e., \textit{near user} in conventional downlink NOMA) can be given in the form
\begin{equation}
P_{x_1}^{(sr)}(e|_{\gamma_{sr}})=\sum_{i=1}^L\eta_iQ\left(\sqrt{2\vartheta_i\rho_s\gamma_{sr}}\right),
\end{equation}
where $L$, $\eta_i$ and $\vartheta_i$ change according to the modulation pairs. For BPSK, when we repeat the steps \cite[Eq.(6)-(9)]{Kara2019}, we obtain $L=5$, $\eta=\sfrac{1}{2}[2, -1, 1, 1, -1]$ and $\vartheta=[\alpha, 1+2\sqrt{\alpha-\alpha^2}, 1-2\sqrt{\alpha-\alpha^2}, 4-3\alpha+4\sqrt{\alpha-\alpha^2},4-3\alpha-4\sqrt{\alpha-\alpha^2}]$.
\begin{IEEEproof}
In order to detect $x_1$ symbols, the relay should firstly detect $x_2$ symbols and subtract these detected $\hat{x}_2$ symbols from the received signal $y_{sr}$. Thus, in the analysis, both correct and erroneous SIC of $x_2$ symbols should be considered. Let us firstly to analyze $b_1b_2=00$ scenario in the correct SIC case. According to Table I, the $y_{sr}=\left(-\sqrt{\alpha}-\sqrt{\left(1-\alpha\right)}\right)h_{sr}+n_{sr}<0$ is satisfied. After the SIC, to detect $x_1$ symbol correctly, for the remaining signal, it should be $y_{sr}^{'}=-\sqrt{\alpha}h_{sr}+n_{sr}<0$. Nevertheless, with the priori probability (i.e., correct SIC), the erroneous detection probability in this case is given by $P\left(n_{sr}<\left(\sqrt{\alpha}+\sqrt{\left(1-\alpha\right)}\right)h_\lambda\right)\times P\left(n_{sr}\geq\sqrt{\alpha}h_{sr}|_{n_{sr}<\left(\sqrt{\alpha}+\sqrt{\left(1-\alpha\right)}\right)h_\lambda}\right)$. By applying the conditional probability rule, it becomes $P\left(\sqrt{\alpha}\leq{n_{sr}<\left(\sqrt{\alpha}+\sqrt{\left(1-\alpha\right)}\right)h_\lambda}\right)$. In the same way, for the erroneous SIC, for $b_1b_2=01$, we know that $y_{sr}=\left(-\sqrt{\alpha}+\sqrt{\left(1-\alpha\right)}\right)h_{sr}+n_{sr}<0$. To detect $x_1$ symbol correctly, the remaining signal (i.e., $y_{sr}^{'}$) should be $y_{sr}^{'}=\left(-\sqrt{\alpha}+2\sqrt{\left(1-\alpha\right)}\right)h_{sr}+n_{sr}<0$. Considering the priori erroneous SIC condition and the decision rule in this case, the conditional error probability for this scenario is obtained as $P\left(\sqrt{\alpha}-2\sqrt{\left(1-\alpha\right)}<{n_{sr}\leq\left(\sqrt{\alpha}-\sqrt{\left(1-\alpha\right)}\right)h_\lambda}\right)$. After repeating the correct and erroneous SIC scenarios for each bit-stream, with some algebraic simplifications, we obtain given $L$, $\eta$, and $\vartheta$ coefficients, so the proof is completed.
\end{IEEEproof}

Then, by averaging (8) over instantaneous $\gamma_{sr}$, just like (5) and (7), the BEP of $x_1$ symbols in the first phase is derived as
\begin{equation*}
\begin{split}
  &P_{x_1}^{(sr)}(e)=\\
  &\begin{cases}
  \sum\limits_{k=1}^5\frac{\eta_k}{2}\left[1-\mu^2(c_k)\sum\limits_{l=0}^{m_{sr}-1}\binom{2l}{l}\left(\frac{1-\mu^2(c_k)}{4}\right)^l\right], m_{sr}:  \text{integer}, \\
  \sum\limits_{k=1}^5\frac{\eta_k}{2\sqrt{\pi}}\frac{\sqrt{c_k}}{\left(1+c_k\right)^{m_{sr}+0.5}}\frac{\Gamma(m_{sr}+0.5)}{\Gamma(m_{sr}+1)} \times\\
  {_2}F_1\left(1,m_{sr}+0.5;m_{sr}+1;\frac{1}{1+c_k}\right),  m_{sr}:  \text{non-integer},
    \end{cases}
    \end{split}\tag9
 \end{equation*} \stepcounter{equation}
where $c_k\triangleq\frac{\vartheta_k\rho_s\Omega_{sr}}{m_{sr}}$.
\begin{figure*}[b]
\centering
\hrulefill
\begin{equation}
P^{(e2e)}\left(e\right)=\begin{cases}
 \sum\limits_{k=1}^2\frac{1}{8}\left[1-\mu^2(b_k)\sum\limits_{l=0}^{m_{sd}-1}\binom{2l}{l}\left(\frac{1-\mu^2(b_k)}{4}\right)^l\right]+
 \sum\limits_{k=1}^5\frac{\eta_k}{4}\left[1-\mu^2(c_k)\sum\limits_{l=0}^{m_{sr}-1}\binom{2l}{l}\left(\frac{1-\mu^2(c_k)}{4}\right)^l\right] \\
 \times\left[1-\frac{1}{2}\left[1-\mu^2(p)\sum\limits_{l=0}^{m_{rd}-1}\binom{2l}{l}\left(\frac{1-\mu^2(p)}{4}\right)^l\right]\right]+
 \left[1-\sum\limits_{k=1}^5\frac{\eta_k}{4}\left[1-\mu^2(c_k)\sum\limits_{l=0}^{m_{sr}-1}\binom{2l}{l}\left(\frac{1-\mu^2(c_k)}{4}\right)^l\right]\right]\\ \times\frac{1}{4}\left[1-\mu^2(p)\sum\limits_{l=0}^{m_{rd}-1}\binom{2l}{l}\left(\frac{1-\mu^2(p)}{4}\right)^l\right], \quad \quad  \quad  \quad  \quad  \quad \quad \quad  \quad  \quad  \quad  \quad \quad \quad  \quad  \quad  \quad  \quad  \forall\lambda, \ m_\lambda:\text{integer}, \\[17pt]
 \sum\limits_{k=1}^2\frac{1}{8\sqrt{\pi}}\frac{\sqrt{b_k}}{\left(1+b_k\right)^{m_{sd}+0.5}}\frac{\Gamma(m_{sd}+0.5)}{\Gamma(m_{sd}+1)} \times
  {_2}F_1\left(1,m_{sd}+0.5;m_{sd}+1;\frac{1}{1+b_k}\right) +\sum\limits_{k=1}^5\frac{\eta_k}{4\sqrt{\pi}}\frac{\sqrt{c_k}}{\left(1+c_k\right)^{m_{sr}+0.5}}\frac{\Gamma(m_{sr}+0.5)}{\Gamma(m_{sr}+1)}\\ \times {_2}F_1\left(1,m_{sr}+0.5;m_{sr}+1;\frac{1}{1+c_k}\right)
  \times \left[1-\frac{1}{2\sqrt{\pi}}\frac{\sqrt{p}}{\left(1+p\right)^{m_{rd}+0.5}}\frac{\Gamma(m_{rd}+0.5)}{\Gamma(m_{rd}+1)} \times {_2}F_1\left(1,m_{rd}+0.5;m_{rd}+1;\frac{1}{1+p}\right)\right]\\
  +\left[1-  \sum\limits_{k=1}^5\frac{\eta_k}{2\sqrt{\pi}}\frac{\sqrt{c_k}}{\left(1+c_k\right)^{m_{sr}+0.5}}\frac{\Gamma(m_{sr}+0.5)}{\Gamma(m_{sr}+1)} \times {_2}F_1\left(1,m_{sr}+0.5;m_{sr}+1;\frac{1}{1+c_k}\right)\right]\frac{1}{4\sqrt{\pi}}\frac{\sqrt{p}}{\left(1+p\right)^{m_{rd}+0.5}}\frac{\Gamma(m_{rd}+0.5)}{\Gamma(m_{rd}+1)}\\ \times{_2}F_1\left(1,m_{rd}+0.5;m_{rd}+1;\frac{1}{1+p}\right),  \quad \quad  \quad  \quad  \quad  \quad \quad \quad  \quad  \quad  \quad  \quad \quad \quad  \quad  \quad  \quad  \quad \text{otherwise}.
 \end{cases}
\end{equation}
\end{figure*}
Finally, by substituting (7) and (9) into (6) and then by substituting (5) and (6) into (3), the ABEP of the NOMA-CRS is derived in the closed-form as (10) (see the bottom of the page). As seen in (10), the ABEP of the NOMA-CRS is presented in a simpler form when all $m_\lambda$ has integer values. 


\vspace{-0.5\baselineskip}
\section{Power Optimization for MBER}
Considering the total power consumption at the nodes, let us define a PS coefficient (i.e., $\beta$). The total power ($P_T$) is shared as $P_s=\beta P_T$ and $P_r=\left(1-\beta\right) P_T$.
The joint PS-PA optimization is defined as the $\alpha, \beta$ pair which minimizes the ABEP of the NOMA-CRS. Thus, it is given as
\begin{equation}
[\alpha^*, \beta^*]=\argmin_{\alpha,\beta}{P^{(e2e)}(e)}.
\end{equation}
And, it is derived by solving
\begin{equation}
[\alpha^*, \beta^*]=\arg\left(\frac{\partial ^2P^{(e2e)}(e)}{\partial\alpha \partial\beta}=0\right).
\end{equation}
To the best of the authors' knowledge, (12) cannot be analytically solved in closed-form. It can be obtained by iterative algorithms such as a brute-force/full-search algorithm \cite{book_IS} by computing the ABEP values for all PS-PA coefficients to obtain the minimum. However, this costs a high computational complexity in the online implementation thereby increasing latency and is inappropriate for practical implementations. Besides, with this high computational operation, high power consumption will be also required at the relay which yields an unfairness for the relay.
Thus, we propose an ML-based model to obtain the optimum PS-PA pair for any condition in NOMA-CRS. The proposed ML network is trained offline to reduce computation time/complexity/latency and it is implemented online as being in all ML-based solutions in physical layer communications \cite{Zappone2019,ML_phy}. Therefore, the computational complexity for solving the optimization problem is shifted to the offline training stage and a very low online implementation complexity is achieved.
\subsection{ML-aided Optimum PS-PA}

\subsubsection{Proposed ML Model}
We built a three-layered (i.e., input, hidden, and output layers) fully connected neural network (NN) model to compute the optimum PS-PA pair for the MBER in NOMA-CRS.  
Hence, we redefine the PS-PA optimization problem
\begin{equation}
[\alpha^*, \beta^*]=f\left(m_\lambda, \Omega_\lambda \right).
\end{equation}
According to (13), the built NN model has 6 inputs (neurons) and 2 outputs (neurons). The number of neurons in hidden layer is $10$ and \textit{Levenberg-Marquardt} is used for the learning algorithm. Minimum performance gradient, learning (mu) decrease factor and learning (mu) increase factor are set as $1e-7$, $0.1$ and $10$, respectively.
The network parameters are empirically determined such that increasing number of layers and/or neurons do not provide a remarkable performance gain and the training performance converges. Therefore, not to increase the online implementation complexity, we came up with that three-layered NN is enough to solve the problem defined in (11)  although it can be built with much deeper networks (more hidden layers).


\subsubsection{Training and Testing}
To train NN model, we firstly create a dataset for different channel and power conditions. Then, we train the NN model to predict the optimum PS-PA pair for given training inputs with minimum MSE compared to desired training outputs. The dataset generation and training algorithm is given in Algorithm 1.
\begin{algorithm}[h]
\caption{Dataset Generation and Training}\label{PA}
\begin{algorithmic}[1]
\State Data Set Generation \\
\textbf {Multiple For Loops} ($\Omega_\lambda=[0:1:10]$, $m_\lambda=[0.5:0.5:4]$ and $\rho_T=P_T/N_0=[0:5:20]$ dB)
\State Solve (12) by numerical tools, then, label the obtained results as desired $[\alpha^*,\beta^*]$ outputs for the given inputs
\State \textbf {End For Loops}
\State Portion the dataset $90\%$ for training and $10\%$ for testing
\State \textbf {While}
\State Train the network with the training inputs ($90\%$) to minimize the MSE
\State Test the trained network for the testing inputs ($10\%$)
\State \textbf{Unless} the test performance satisfies
\end{algorithmic}
\end{algorithm}
In Algorithm 1, we divide the dataset into two groups: $90\%$ for training and $10\%$ for test datasets. We train the network with training dataset to minimize the MSE where we use a cease criterion that we stop training if the MSE is not improved at least by $e-5$ within two consecutive epochs. Then, we test the trained network with the test dataset (i.e., which is not used in training stage) and re-train the network until we obtain a good (convincing) test performance. At the end of this re-training procedure, we have concluded with the best performance metrics as obtained $3.43e-4$ MSE and $0.9984$ and regression for training and $3.44e-4$ MSE and $0.9842$ regression for testing compared to desired (numerically calculated) optimum PS-PA pairs.
\vspace{-1\baselineskip}
\subsection{Complexity}
In the ML-based solutions, the computational complexity is related to the online implementation. Therefore, we focus on the online implementation complexity (feed-forward calculation). The online implementation complexity is just $\mathcal{O}\left(104\right)$ which consists of $60$ and $20$ weight multiplications from-input-to-hidden and from-hidden-to-output layers, respectively. Total $12$ adds for biases and $12$ activation functions (i.e., $tansig()$ function) are computed on hidden and output layers neurons. On the other hand, to optimize PS-PA by an iterative algorithm (i.e., brute-force/full-search algorithm), the ABEP expression (10) should be computed for all PA-PS pairs and compared with each other. Thus, by considering arithmetic/logical operations in (10) (includes computing high-complex $\Gamma(.)$ and ${_2}F_1\left(,;;\right)$ functions), the complexity of the full search algorithm is obtained as $\mathcal{O}\left(M^2\left(2m_{sd}+5m_{sr}+m_{rd}+96\right)+1\right)$ \cite{book_IS} for non-integer case where $\sfrac{1}{M}$ denotes the step size
(resolution in search) for PS and PA. For even all integer case, it reduces to  $\mathcal{O}\left(2M^2\left(2m_{sd}+5m_{sr}+m_{rd}+9\right)+1\right)$. Besides, this computational complexity should be repeated whenever the channel conditions change. However, the proposed model has only $\mathcal{O}\left(104\right)$ online complexity. Therefore, the complexity (a.k.a. latency) is reduced, and the optimization is implementable in practical scenarios. Since we propose a joint PS-PA optimization, this should be computed at both source and relay. Therefore, reducing the complexity is essential. With the proposed method, we shifted the complexity to the offline training stage (at a server with high computational capacity); thus, the online implementation complexity at the source and relay is limited (e.g., centralized offline training and distributed online implementation). This is also very important for extended scenarios such as multi-relay schemes where the same online complexity will be required even if a relay selection is applied \cite{Kara2016}.

\begin{figure*}[ht]
\centering
\subfloat[$m_\lambda=$ $\left\{0.5,1,1.5,2\right\}$,$ \Omega_{sr}=\Omega_{rd}=2, \Omega_{sd}=1$ and $ \Omega_{sr}=\Omega_{rd}=10, \Omega_{sd}=2$ ]{\includegraphics[width=6cm]{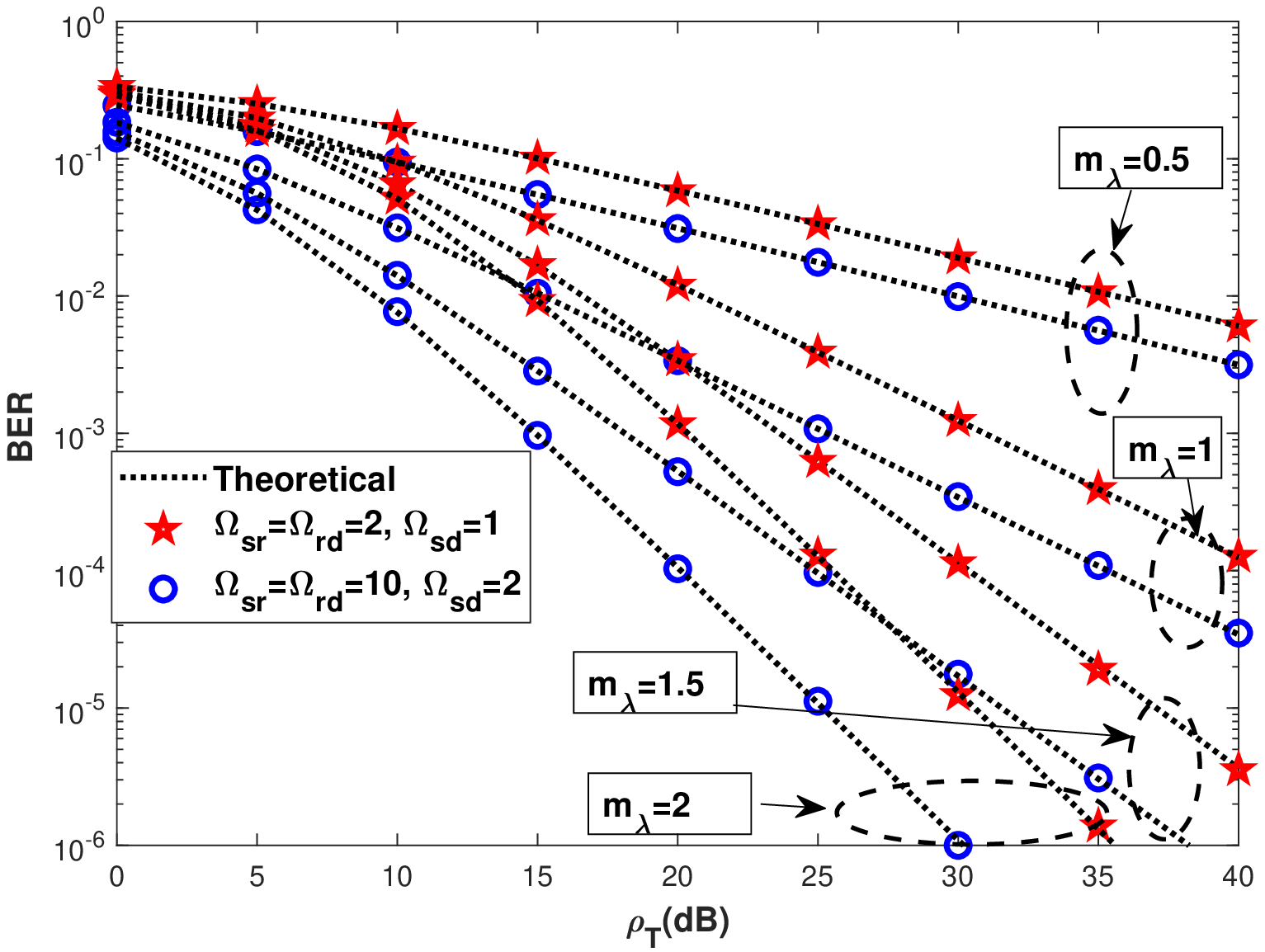}
\label{ber1}}
\subfloat[$ \Omega_{sr}=\Omega_{rd}=2,\Omega_{sd}=1$]{\includegraphics[width=6cm]{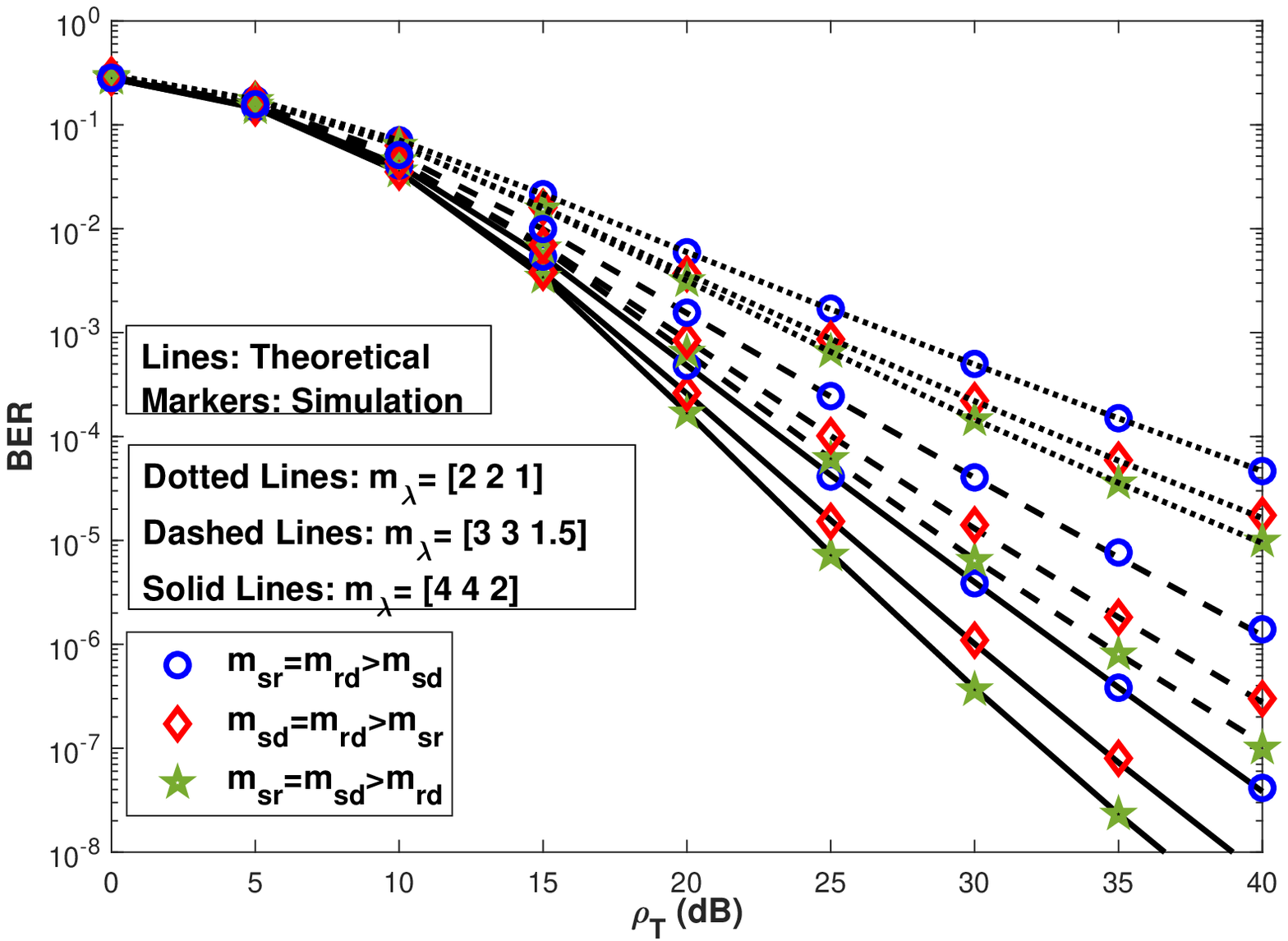}
\label{ber2}}
\subfloat[Comparisons for fixed, full-search and proposed PS-PA schemes]{\includegraphics[width=6cm]{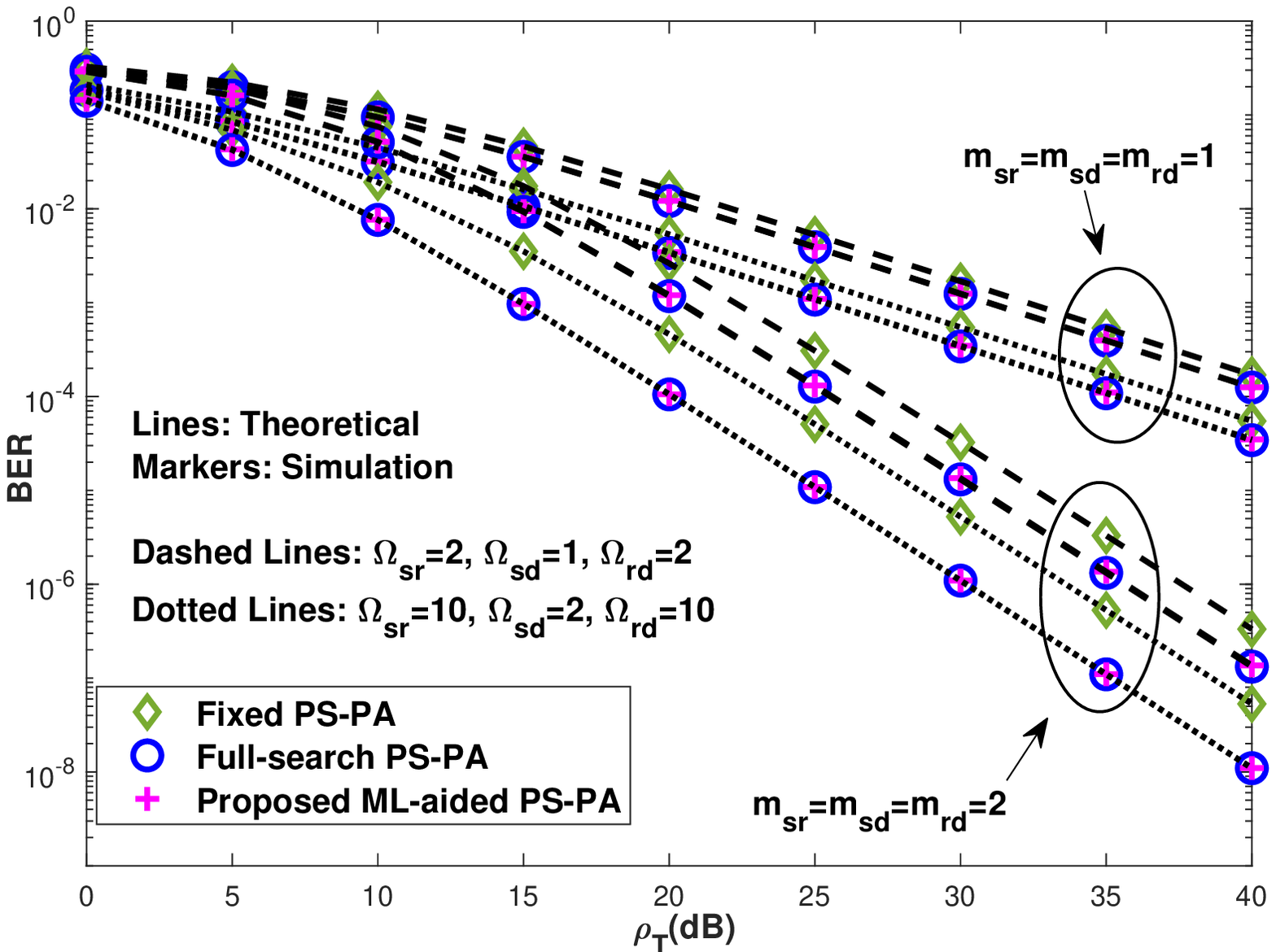}
\label{diversity}}
\caption{BER of the NOMA-CRS vs SNR.}
\label{vs_SNR}
\end{figure*}


\section{Numerical Results}
The Fig. 1 presents the BER of the NOMA-CRS for different shape and spread parameters with respect to $\rho_T$. In Fig. 1.a, it is assumed that $m_\lambda$ is equal $\forall \lambda$. The outcomes of the online implementation of the proposed ML network are used as PS-PA strategy. It is noteworthy that the theoretical analysis matches perfectly with the simulations. As expected, the shape parameter ($m_\lambda$) denotes the diversity order (superscript of the $\sfrac{1}{SNR}$ when $SNR\rightarrow\infty$) while increasing the spread parameter ($m_\lambda$) provides only a horizontal gain in error performance. Then, to reveal the effect of the shape parameter for each node, we present the BER of the NOMA-CRS in Fig. 1.b when $m_\lambda$ is not equal $\forall \lambda$. Likewise in conventional CRS, the diversity order of the NOMA-CRS is driven by the lowest shape parameter (i.e., $\min\{m_{sr},m_{sd},m_{rd}\}$). Nevertheless, due to the cooperative communication, if an error propagation occurs from the relay to the destination, the full diversity cannot be observed. Hence, to achieve the full diversity, by considering the total consumed power, the PS and PA should be jointly optimized for MBER. For instance, when $m_{sr}=m_{sd}=4$ and $m_{rd}=2$, the diversity order of the NOMA-CRS is equal to 2. Nevertheless, the other $m$ parameters have also an effect on the error performance, especially in the high SNR region. With lower $m$ parameter between S-R (e.g., $m_{sd}=m_{rd}=4, m_{sr}=2$), the $x_1$ symbols in the first phase has a poor error performance, thus it causes an error propagation from relay-to-source and the error performance gets worse. It can be easily seen in high SNR regime although this case has also the diversity order of 2.
Since the total BER is the average of two symbols, the BER of NOMA-CRS highly depends on their individual BEPs. Hence, we should guarantee that none of them pulls down the BER.
Then, to reveal the effectiveness of the proposed PS-PA scheme, we present the BER of the NOMA-CRS for the proposed ML-aided PS-PA, fixed PS-PA, and full-search PS-PA in Fig. 1.c. The fixed PS-PA strategy is $\beta=0.5$ (i.e., $P_s=P_r$) and $\alpha=0.2$ as assumed in previous NOMA-CRS studies \cite{Kim2015,Jiao2017,Xu2016,Zhang2018,Wan2019,Abbasi2019}. One can easily see that the proposed ML-aided PS-PA performs the same with the full search PS-PA and it outperforms the fixed PS-PA significantly. Hereby, we should note that the online complexity of the proposed ML-aided PS-PA is much less than the full search PS-PA as proved in Section IV.B. In addition, this performance gain over the fixed PS-PA becomes greater with the increase of shape ($m_\lambda$) and/or spread ($\Omega_\lambda$) parameters. 

We present PS-PA comparisons between the outcomes of the proposed NN network and the results obtained by the full search algorithm for various scenarios in Fig. 2. The predictions of the proposed NN network for PS-PA scheme are very close to the full search algorithm. Furthermore, the PS-PA scheme is highly dependant on the relay position. To represent this, we hereby assume that the linear sum of the spread parameters through the link S-R-D is constant (i.e., $\Omega_{rd}+\Omega_{sr}=10$), thus $\Omega_{rd}=10-\Omega_{sr}$ which can call the relay position (when $\Omega_{sr}>\Omega_{rd}$, relay is close to source vice versa). It is clearly seen that the optimum PS-PA values get larger when the channel quality between S-R (i.e, $\Omega_{sr}$) decreases. It can be explained as follows. If the relay detects $x_1$ symbol erroneously, an error propagation occurs from the relay to the destination and this pulls down the total error performance. Hence, not to cause an error propagation, most of the power should be transferred to $x_1$ symbols (higher $\alpha$ and $\beta$), However, increasing $\alpha$ and/or $\beta$ too much still causes an error propagation due to the SIC operation at the relay. Thus, they should be limited.

\begin{figure}[ht]
		\centering
    \includegraphics[width=7cm]{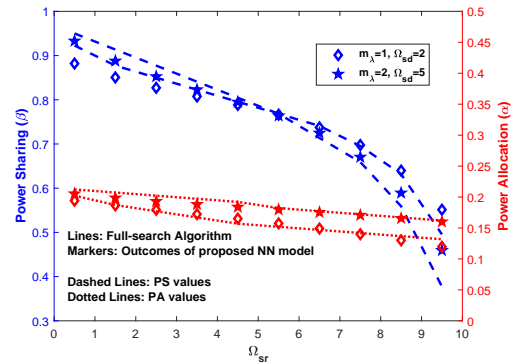}
    \caption{Optimum PS-PA for NOMA-CRS under MBER constraint.}
    \label{opt_PA}
\end{figure}
Fig. 3 presents BER comparisons between the proposed PS-PA and the optimum PA (in terms of capacity) in \cite{Kim2015} with the fixed PS. In Fig. 3.a, since a joint PS-PA is not implemented in \cite{Kim2015}, an error propagation occurs in the high SNR region, whereas with the proposed PS-PA, this problem has been revolved. Besides, to emphasize the effect of PS-PA, Fig. 3.b shows the error performance of the NOMA-CRS with respect to $\alpha$ and $\beta$. NOMA-CRS has the best performance when $0.1\leq\alpha\leq 0.2$ and $0.7\leq\beta\leq 0.9$. Increasing/decreasing one/both of them too much causes severe performance for $x_1$ and/or $x_2$ symbols so for the NOMA-CRS. In Fig 3.b, we also mark the points when the proposed PS-PA and the PA in \cite{Kim2015} are used. In both figures, it is clear that the proposed ML-aided optimum PS-PA provides the minimum BER for NOMA-CRS and is superior to the other PA strategies.
\begin{figure}[ht]
		\centering
    \includegraphics[width=8.5cm]{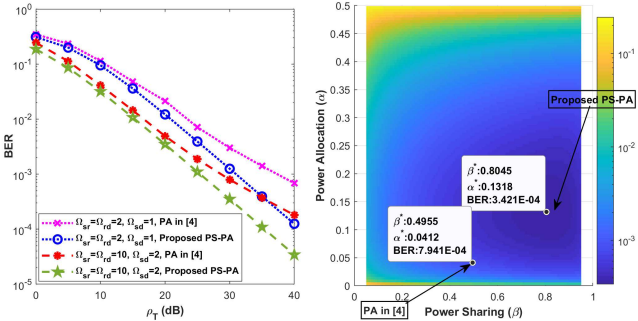}
    \caption{BER of the NOMA-CRS for $m_\lambda=1$ (Rayleigh), $\Omega_{sr}=\Omega_{rd}=10$ and $\Omega_{sd}=2$ a) vs. SNR b) vs. $\alpha$ and  $\beta$ when $\rho_s=30$ dB.}
    \label{ber_PA}
\end{figure}

Lastly, to represent the effectiveness of the proposed PS-PA for higher modulation order schemes, we present BER comparisons for the fixed PS-PA (i.e., $\beta=0.5$, $\alpha=0.2$) \cite{Kim2015,Jiao2017,Xu2016,Zhang2018,Wan2019,Abbasi2019}, full-search PS-PA and the proposed ML-aided PS-PA in Fig. 4 when QPSK is used for both symbols. Since, the theoretical analysis for QPSK has not been derived, yet, the full-search PS-PA is obtained by simulations where we simulate the NOMA-CRS for $100\times100$ PS-PA pairs and chose the PS-PA which has minimum BER. As seen in Fig. 4, the proposed ML-aided PS-PA still outperforms the fixed PS-PA and achieves the full-search PS-PA performance. This simulations can be extended for higher M-QAM schemes; however, in M-QAM signaling, we should define an additional PA constraint for a detectable signal design \cite{Kara2020}.
\begin{figure}[ht]
		\centering
    \includegraphics[width=7cm]{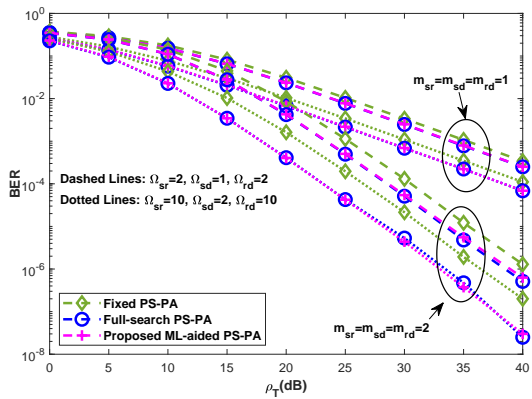}
    \caption{BER comparisons for NOMA-CRS with QPSK}
    \label{opt_PA2}
\end{figure}
\vspace{-1.4\baselineskip}
\section{Conclusion}
In this paper, we investigate the error performance of the NOMA-CRS considering imperfect SIC and derive the exact ABEP of the NOMA-CRS over Nakagami-m fading channels. Then, to minimize  the BER of the NOMA-CRS, we propose an ML-aided optimum PS-PA strategy. Based on extensive simulations, our proposed ML-aided PS-PA strategy is optimal in terms of MBER criteria and provides a remarkable gain compared to existing PA strategies. This study is the first to optimize PS-PA in NOMA schemes under the BER constraint; thus, it can be further extended for other NOMA systems by re-training the proposed network with the dataset of that NOMA scheme which is seen as a future work.



%



\ifCLASSOPTIONcaptionsoff
  \newpage
\fi



%
\bibliographystyle{IEEEtran}
\bibliography{kara_VT_2021_00627}

%






\end{document}